\documentclass{aastex}\newcommand{\emul}{true}

\newcommand{\IncludeFigures}{true}

\usepackage{amsmath}
\usepackage{amssymb}
\usepackage{xspace}

\usepackage{ifthen}
\newboolean{emul}%
\setboolean{emul}{\emul}%

\ifthenelse{\boolean{emul}}{
  \usepackage{emulateapj5}
  \usepackage{apjfonts}
  \setcounter{secnumdepth}{5}
  \usepackage{graphicx}
  \setcounter{topnumber}{3}
  \setcounter{bottomnumber}{2}
  \setcounter{totalnumber}{5}
  \setcounter{dbltopnumber}{2}

  \topmargin 0.5in
  \submitted{Draft \today}
}{}

\lefthead{BARAFFE, HEGER, AND WOOSLEY}
\righthead{}

\newcommand{\powersep}{{\ensuremath{\times}}}

\newcommand{\g}{{\ensuremath{\mathrm{g}}}\xspace}
\newcommand{\K}{{\ensuremath{\mathrm{K}}}\xspace}
\newcommand{\cm}{{\ensuremath{\mathrm{cm}}}\xspace}
\newcommand{\yr}{{\ensuremath{\mathrm{yr}}}\xspace}

\newcommand{\Day}{{\ensuremath{\mathrm{d}}}\xspace}
\newcommand{\Msun}{{\ensuremath{\mathrm{M}_{\odot}}}\xspace}

\newcommand{\Rsun}{{\ensuremath{\mathrm{R}_{\odot}}}\xspace}
\newcommand{\Sec}{{\ensuremath{\mathrm{s}}}\xspace}
\newcommand{\erg}{{\ensuremath{\mathrm{erg}}}\xspace}

\newcommand{\cms}{{\ensuremath{\cm\,\Sec^{-1}}}\xspace}

\newcommand{\scmg}{{\ensuremath{\cm^2\,\g^{-1}}}\xspace}

\newcommand{\erggs}{{\ensuremath{\erg\,\g^{-1}\,\Sec^{-1}}}\xspace}

\newcommand{\img}{{\ensuremath{\mathbf{i}}}\xspace}

\newcommand{\Msunyr}{{\ensuremath{\Msun\,\yr^{-1}}}\xspace}
\newcommand{\kK}{{\ensuremath{\mathrm{k}\K}}\xspace}
\newcommand{\kyr}{{\ensuremath{\mathrm{k}\yr}}\xspace}
\newcommand{\Myr}{{\ensuremath{\mathrm{M}\yr}}\xspace}

\newcommand{\Mdot}{{\ensuremath{\dot{M}}}\xspace}
\newcommand{\Tc}{{\ensuremath{T_{\mathrm{c}}}}\xspace}

\newcommand{\Teff}{{\ensuremath{T_{\mathrm{eff}}}}\xspace}
\newcommand{\taud}{{\ensuremath{\tau_{\mathrm{d}}}}\xspace}
\newcommand{\Amax}{{\ensuremath{A_{\mathrm{max}}}}\xspace}
\newcommand{\tH}{{\ensuremath{\tau_{\mathrm{H}}}}\xspace}
\newcommand{\tS}{{\ensuremath{\tau_{\mathrm{stable}}}}\xspace}

\newcommand{\Fconv}{{\ensuremath{F_{\mathrm{conv}}}}\xspace}

\newcommand{\Yc}{{\ensuremath{Y_{\mathrm{c}}}}\xspace}
\newcommand{\enuc}{{\ensuremath{\epsilon_{\mathrm{nuc}}}}\xspace}
\newcommand{\vesc}{{\ensuremath{v_{\mathrm{esc}}}}\xspace}

\newcommand{\dr}{{\ensuremath{\delta\!r}}\xspace}

\newcommand{\drs}{{\ensuremath{\dr_{\!\mathrm{surf}}}}\xspace}
\newcommand{\DM}{{\ensuremath{\Delta\!M}}\xspace}
\newcommand{\EP}{{\ensuremath{E_{\mathrm{puls}}}}\xspace}
\newcommand{\csound}{{\ensuremath{c_{\mathrm{sound}}}}\xspace}

\newcommand{\Lpuls}{{\ensuremath{L_{\mathrm{puls}}}}\xspace}

\newcommand{\lSect}[1]{{\label{sec:#1}}}
\newcommand{\lFig}[1]{{\label{fig:#1}}}
\newcommand{\lEq}[1]{{\label{eq:#1}}}
\newcommand{\lTab}[1]{{\label{tab:#1}}}
\newcommand{\pFig}[1]{{\placefigure{fig:#1}}}

\newcommand{\Tabff}[1]{{\ref{tab:#1}}}
\newcommand{\Tab}[1]{{Table~\Tabff{#1}}}

\newcommand{\pan}[1]{{\textit{#1}}}

\newcommand{\FIGFF}[2]{{\ref{fig:#2}\pan{#1}}}

\newcommand{\FIG}[2]{{Fig.~\FIGFF{#1}{#2}}}
\newcommand{\Fig}[2][]{{\FIG{#1}{#2}}}

\newcommand{\Figure}[1]{{Figure~\FIGFF{}{#1}}}

\newcommand{\Sectff}[1]{{\ref{sec:#1}}}
\newcommand{\Sect}[1]{{Sect.~\Sectff{#1}}}

\newcommand{\Eqref}[1]{{\ref{eq:#1}}}
\newcommand{\Eqff}[1]{{(\Eqref{#1})}}
\newcommand{\Eq}[1]{{Eq.~\Eqff{#1}}}
\newcommand{\Eqs}[1]{{Eqs.~\Eqff{#1}}}

\newcommand{\isofont}[1]{{\mathrm{#1}}}
\newcommand{\isomass}[1]{{\ensuremath{\isofont{^{#1}}}}}
\newcommand{\isocharge}[1]{{\ensuremath{\isofont{_{#1}}}}}
\newcommand{\isotope}[3]{{\ensuremath{\isocharge{#1}\isomass{#2}\isofont{#3}}}}

\newcommand{\I}[2]{{\isotope{}{#1}{#2}}}

\newcommand{\Ep}[1]{{\ensuremath{10^{#1}}}}
\newcommand{\E}[1]{{\ensuremath{\powersep\Ep{#1}}}}

\newcommand{\msol}{\Msun}

\newcommand{\te}{\Teff}

\newcommand{\simgr}{{\ensuremath{\gtrsim}}\xspace}
\newcommand{\simle}{{\ensuremath{\lesssim}}\xspace}

\ifthenelse{\boolean{emul}}{\renewcommand{\pFig}[1]{}}

\newcommand{\FigHRDmassFile}{fig1.ps} 
\newcommand{\FigHRDmass}{%
Theoretical H-R diagram for stars with $Z=0$ and different masses
assuming no mass loss.  The time-scales of evolution are indicated by
marks on the curves.  \lFig{Z0-HRD}}

\newcommand{\FigHRDZFile}{fig2.ps}
\newcommand{\FigHRDZ}{%
Theoretical HR diagram for 120\,\Msun stars with different $Z$ from
hydrogen ignition till 1\,\% of hydrogen left in the center of the
star -- assuming no mass loss. $[Z]$ corresponds to $\log \, (Z/Z_\odot)$
\lFig{Z-HRD}}

\newcommand{\FigWorkFileA}{fig3_color.ps}
\newcommand{\FigWork}{%
\textbf{(a)} Differential work d$W$/d$M$, in arbitrary units, as a
function of the temperature ($T$ in K) in the interior structure of a 300\,\Msun
with $Z=0$, on the ZAMS.  The Rosseland mean opacity $\kappa$ (in
\scmg) is indicated by the dash-dotted line. The dashed line
corresponds to the nuclear energy generation $\epsilon$ (in units of
5\E5\,\erggs).
\textbf{(b)} Same as (a), but as a function of enclosed mass.
\lFig{dWdm}}

\newcommand{\FigGrowthRateFile}{fig4.ps} \newcommand{\FigGrowthRate}{%
Stability coefficient, $K$, as a function of time.  Positive values
indicate growth of the pulsation amplitude while negative values
(\textit{gray region}) correspond to damped oscillations.  Shown are
120\,\Msun (\textit{solid line}), 200\,\Msun (\textit{dashed}),
300\,\Msun (\textit{dash-dotted}), and 500\,\Msun
(\textit{dash-triple-dotted}) stars of $Z=0$ (assuming no mass loss).
The filled circles indicate the time where these stars become
pulsationally stable.  \lFig{K} }

\newcommand{\FigAmplitudeFile}{fig5.ps} \newcommand{\FigAmplitude}{%
Time integration of the stability coefficient, $K$.  Shown is the
integral $\int_0^{t}K(t')$d$t'$ for 120\,\Msun (\textit{solid line}),
200\,\Msun (\textit{dashed}), 300\,\Msun (\textit{dash-dotted}), and
500\,\Msun (\textit{dash-triple-dotted}) stars of $Z=0$ (assuming no
mass loss).  This gives the number of $e$-folds by which an initial
pulsation would grow till that time.  The filled circles indicate the
time where the stars become pulsationally stable.  After that the
pulsation amplitude is damped quickly.  The slope of the curves is
shown in \Fig{K}.  \lFig{A} }

\ifthenelse{\boolean{emul}}{}{
\slugcomment{Draft for ApJ, \today} 
}

\begin{document}

\title{On the stability of very massive primordial stars}

\author{I. Baraffe}
\affil{Ecole Normale Sup\'erieure, C.R.A.L (UMR 5574 CNRS), 69364 Lyon
Cedex 07, France}
\email{ibaraffe@ens-lyon.fr}

\vskip 0.1 in
\author{A. Heger and S. E. Woosley}
\affil{Department of Astronomy and Astrophysics \\
University of California, Santa Cruz, CA 95064}
\email{alex@ucolick.org, woosley@ucolick.org}
\vskip 0.1 in

\begin{abstract} 

The stability of metal-free very massive stars ($Z$ = 0; $M = 120 -
500\,\msol$) is analyzed and compared with metal-enriched stars. 
Such zero-metal
stars are  unstable to nuclear-powered radial pulsations on the
main sequence, but the growth time scale for these instabilities
is much longer than for their metal-rich counterparts. Since they
stabilize quickly after evolving off the ZAMS, the pulsation may not
have sufficient time to drive appreciable mass loss in Z = 0 stars.
For reasonable assumptions regarding the efficiency of converting
pulsational energy into mass loss, we find that, even for the larger
masses considered, the star may die without losing a large fraction of
its mass.  We find a transition between the $\epsilon$- and
$\kappa$-mechanisms for pulsational instability at $Z\sim
2\E{-4} - 2\E{-3}$.  For the most metal-rich stars, the
$\kappa$-mechanism yields much shorter $e$-folding times, indicating
the presence of a strong instability.  We thus stress the fundamental
difference of the stability and late stages of evolution between very
massive stars born in the early universe and those that might be born
today.

\end{abstract}



\section{Introduction}
\lSect{intro}

The formation and nature of the first generation of so-called
Population III (Pop III) stars have been studied and speculated about
for over thirty years (Schwarzschild \& Spitzer 1953; Yoneyama 1972;
Hartquist \& Cameron 1977; Palla, Salpeter \& Stahler 1983; Silk 1983;
and many others), but the extent to which they differed from present day
stars in ways other than composition is still debated.  Recently,
three-dimensional cosmological simulations have reached sufficient
resolution on small scales to begin to address the star formation
problem (Ostriker \& Gnedin 1996; Abel et al.~1998).  While
considerable uncertainty remains regarding the continued evolution of
the dense knots they find in their calculations, these simulations 
do not exclude the formation of a first generation of
quite  massive  stars 
$\sim 100-1000\,\Msun$ (\textsl{e.g.}, Larson 1999; Abel, Bryan, \& Norman
2000). Using different numerical methods to analyze the fragmentation
of primordial clouds, Bromm et al.~(1999) reached similar conclusions.
Without further study, especially including the complexity of
radiation transport and molecular opacities, it may be premature to
conclude that {\sl all} Pop III stars were so massive, but there
certainly exists adequate motivation to examine the properties of such
stars. In particular, if such massive Pop III stars were ever born,
would they retain their large masses until death? That is, would they
die as pair-instability supernovae, with possible unique signatures of
nucleosynthesis and black hole formation, or would they lose most of
their mass and die much the same as present day stars?

It is well known (Ledoux 1941; Schwarzschild \& H\"arm 1959) that,
above a critical mass, main sequence stars are vibrationally unstable
due to the destabilizing effect of nuclear reactions in their central
regions.  For stars of Pop I composition, the critical mass
for this instability is $\sim 100\, \msol$ (Schwarzschild \& H\"arm
1959).  According to non-linear calculations, such an instability
leads to mass loss rather than catastrophic disruption (Appenzeller
1970a; Talbot 1971ab; Papaloizou 1973ab). Consequently, it has
been thought that stars more massive than about 100 \Msun would not
survive for long before losing a substantial fraction of their
mass. Although the stability of very massive stars has well been
explored since the work of Ledoux (1941), somewhat surprisingly, all
these analyses have been devoted to solar-like compositions or slightly
metal-poor stars (Schwarzschild \& H\"arm 1959; Stothers \& Simon
1968; Aizenman et al.~1975; Stothers 1992; Glatzel \& Kiriakidis 1993;
Kiriakidis et al.~1993), or to helium-core objects such as Wolf-Rayet
stars (Maeder 1985; Cox \& Cahn 1988). Since the temperature
dependence of hydrogen burning and the surface gravitational potential
of low metal stars are quite different, it is important to learn
whether their stability criteria may also vary.

We have thus performed, to our knowledge, the first stability analysis
of zero metallicity stars with $100 \le M/\msol \le 500$ and have also
studied the stability of stars in this mass range
for variable initial metal mass fraction $0 \le Z \le 0.02$.  The
stellar models are presented in \Sect{model}.  The results of the
linear stability analysis are given in \Sect{stalin}, mass loss
estimates are presented in \Sect{Mdot}, and our conclusions follow in
\Sect{concl}.

\section{Stellar evolutionary models}
\lSect{model}

The evolution of four stars of 120, 200, 300 and 500\,\msol was
followed from the Zero Age Main Sequence (ZAMS) to the end of core
helium burning using an implicit Lagrangian hydrodynamics code adapted
to stellar modeling (Langer et al.~1988, Heger 1998; Heger et
al.~2000a).  Details of the code can be found elsewhere (Heger \&
Langer 2000), but to summarize the main physics, the opacity is taken
from Iglesias and Rogers (1996) and the principal nuclear reaction
rates from Caughlan \& Fowler (1988) with the rate for
$^{12}$C($\alpha,\gamma)^{16}$O multiplied by a constant 1.7 (see also
Buchmann 1996).  Mass loss is not taken into account, except for what
is driven by the pulsations studied here.  For zero metallicity stars,
radiative mass loss is probably negligible (Kudritzki 2000).

\ifthenelse{\boolean{emul}}{
\vspace{1.5\baselineskip}
\noindent
\includegraphics[width=\columnwidth]{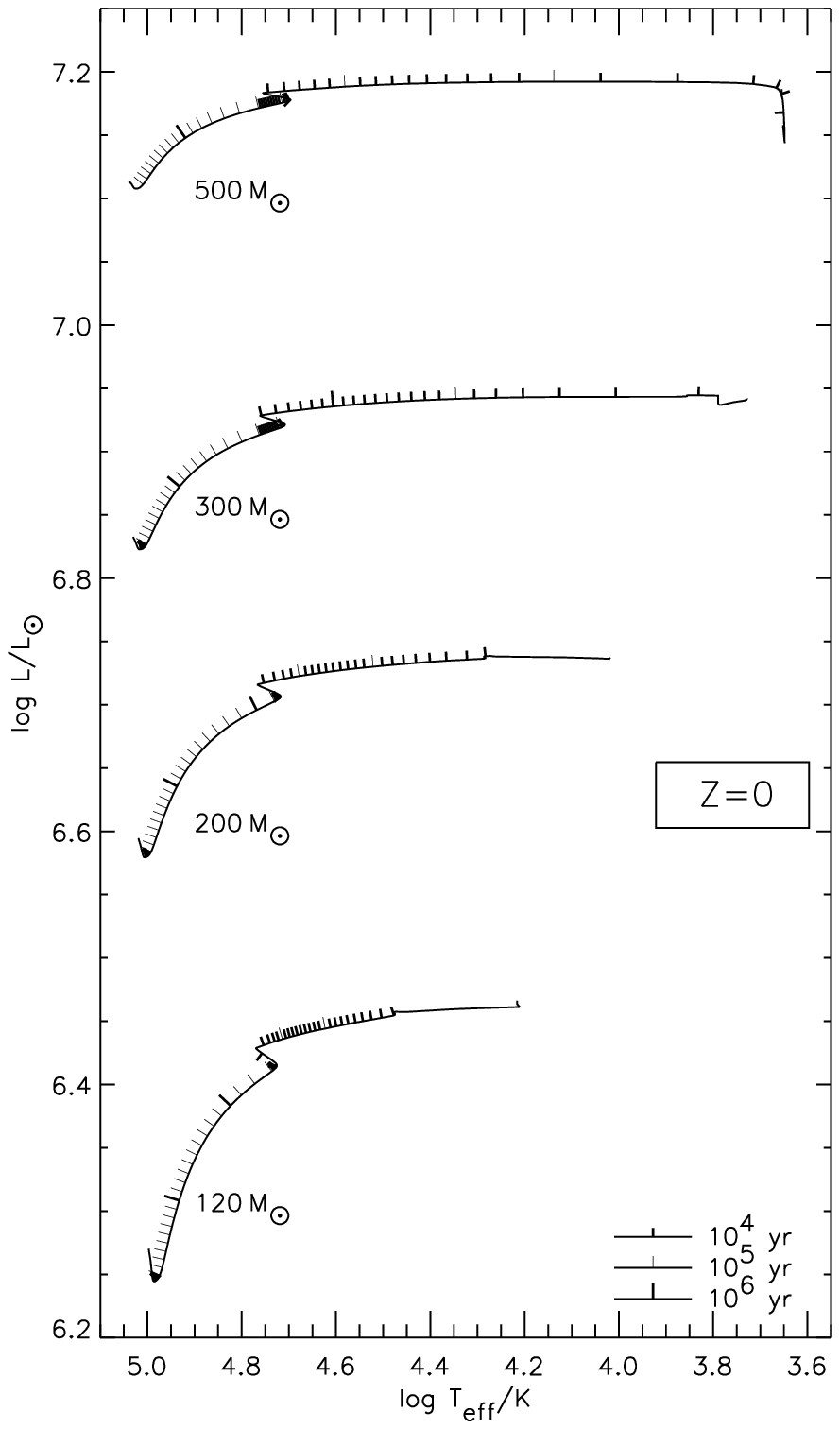}
\figcaption{\FigHRDmass}
\vspace{1.5\baselineskip}}{}

As was first noted by Ezer and Cameron (1971), massive stars with no
metals do not obtain their energy on the main sequence from the
pp-cycle as one might naively expect.  Instead the star contracts to
$\simgr\Ep{8}\,\K$, burns a trace of helium to carbon,
and then expands slightly to burn hydrogen by the CNO cycle.  A CNO
mass fraction of just a few times \Ep{-9} is sufficient to catalyze
even the enormous energy generation of such stars.  However, owing to
this small CNO mass fraction, the star retains a central temperature
of $\sim \Ep8\,\K$ throughout central hydrogen burning.  As Ezer \&
Cameron (1971) also noted, these stars are thus much more compact than
their metal-enriched counterparts and their radiation, while on the
main sequence, is in the ultraviolet (\Tab{ZAMS}).

\Figure{Z0-HRD} shows the theoretical tracks of our $Z=0$ stars in a
Hertzsprung-Russell diagram (HRD).  All begin to contract dynamically
after core helium burning and the star ceases to evolve in the HRD due
to shorter evolutionary time-scale of the core compared to the thermal
time-scale of the envelope.  Note however that, for $M \, \simgr \,
300\,\Msun$, even metal-free stars evolve toward $\te < \Ep4\,\K$ and
eventually become red supergiants, as the hydrogen-burning shell
becomes more active with increasing stellar mass.

In order to analyze the pulsational stability as a function of
metallicity, we repeated our calculations for a wide
range of $Z$ from $Z = 2 \times \Ep{-6}$ to solar, 
$Z = 0.02$, for the 120 $\msol$, and
for selected values of $Z \, > \, 0$ for $M \, > 120 \, \msol$.
 The results are displayed in
\Fig{Z-HRD} for a $120\,\Msun$ star.  Since most of the stability
analyses mentioned in \Sect{intro} (Ledoux 1941; Schwarzschild \&
H\"arm 1959; Appenzeller 1970a; and others) are based upon models with
solar metallicity, but computed using only electron scattering
opacity, we also calculated, for comparison, models with such an opacity
(\Fig{Z-HRD}).

\ifthenelse{\boolean{emul}}{
\vspace{1.5\baselineskip}
\noindent
\includegraphics[width=\columnwidth]{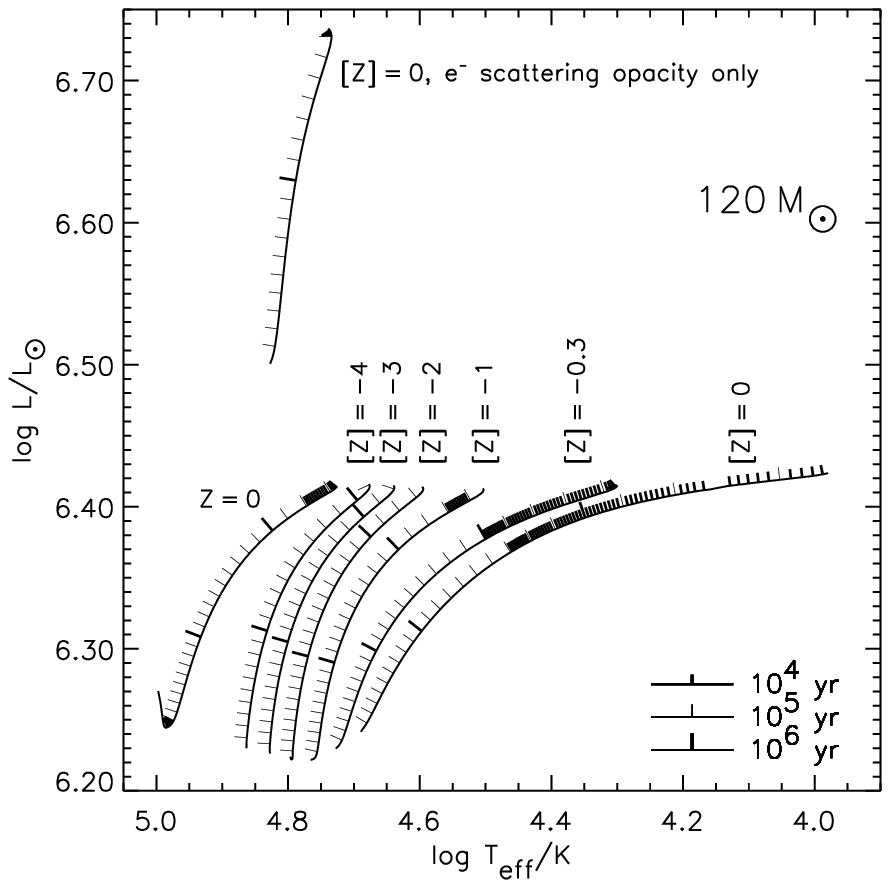}
\figcaption{\FigHRDZ}
\vspace{1.5\baselineskip}}{}

\section{Linear stability analysis}
\lSect{stalin}

A linear non-adiabatic stability analysis of all models was carried
out using a radial pulsation code originally developed by Umin Lee
(1985; see also Heger et al.~1997; Baraffe et al.~1998; Alibert et
al.~1999 for recent applications).  The main uncertainty in these
calculations is the treatment of convection, which is treated in the
``frozen-in'' approximation, \textsl{i.e.}, the perturbation,
$\delta\Fconv$, of the convective flux is neglected in the linearized
energy equation.  Although this approximation is an unavoidable
consequence of our poor knowledge of the interaction between
convection and pulsation, it should not be crucial.  As we shall see,
the relevant regions for pulsation, which are located in
the deep interior of the stars, are fully convective and can be
considered as adiabatic.  In the outer regions, which may also be
important for the pulsation, the convective flux is not the dominant
energy transport.

The stability of the fundamental mode and the overtones (up to the
fifth overtone) was analyzed starting from ZAMS models up to the end of
core helium burning for $Z = 0$ stars. 

\ifthenelse{\boolean{emul}}{
\noindent
\includegraphics[width=\columnwidth]{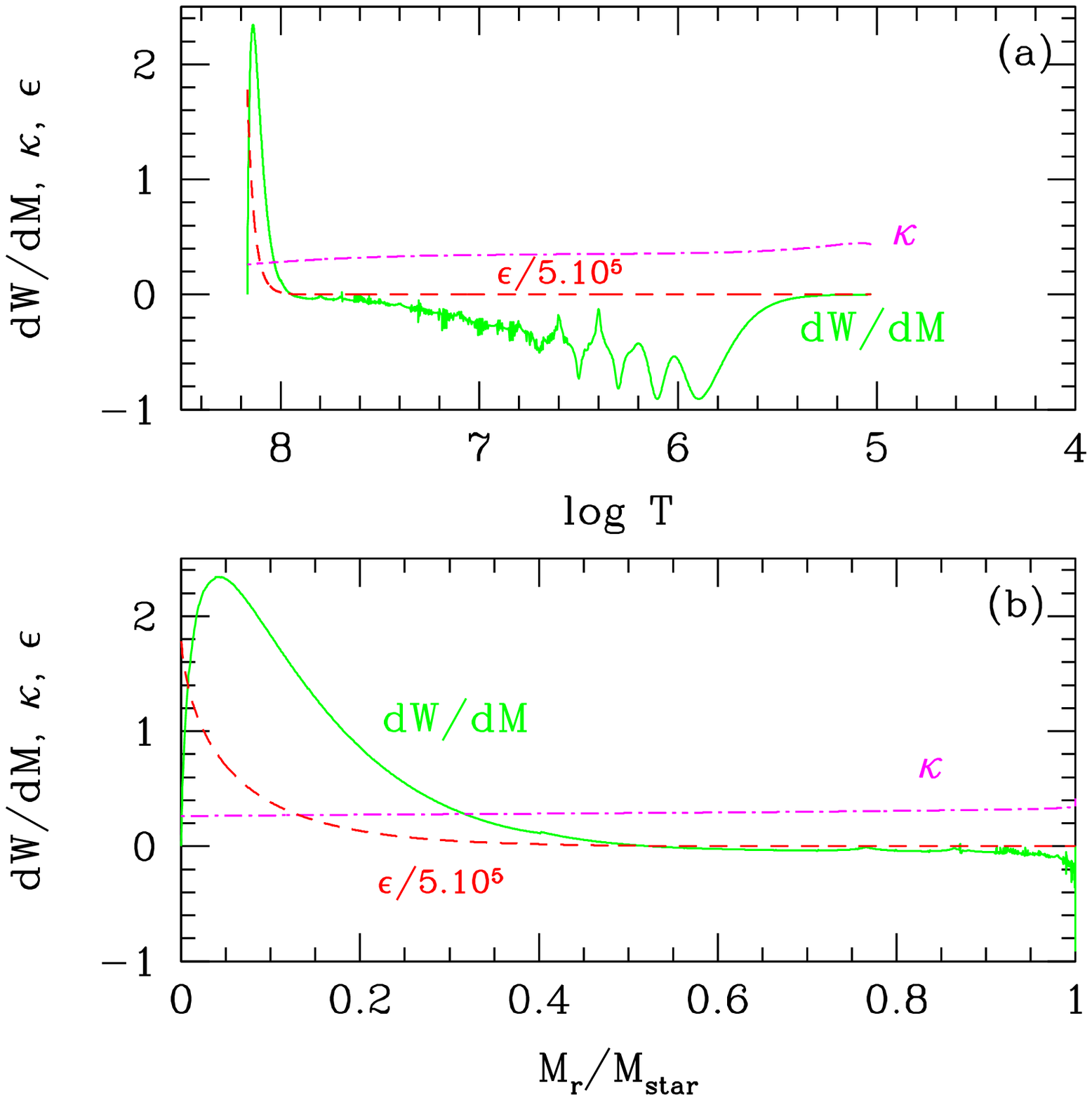}
\figcaption{\FigWork}
}{}

\subsection{Metal-free stars}
\lSect{Z0}

\subsubsection{ZAMS models}
\lSect{ZAMS}

Table 1 gives, for metal-free ZAMS stars, the effective temperature,
\te, stellar radius, $R$, central temperature, \Tc, pulsation period,
$P$, and the pulsational $e$-folding time, \taud (Cox 1980).  Adopting
a time dependence for the eigenfunctions of the form $\exp(\img\sigma
t)\,\exp(Kt)$, where $\sigma = 2\pi/P$ is the eigen-frequency and
$K$, the stability coefficient, the $e$-folding time for the growth of
the instability is given by $\taud = 1/K$.  For all masses studied,
the ZAMS models are unstable and oscillate in the fundamental mode,
whereas the overtones are stable. 
The excitation mechanism is related
to the temperature and density dependence of the nuclear energy
generation rate near the center of the star, the so-called
$\epsilon$-mechanism. As was shown by Ledoux (1941), this driving
mechanism is efficient because of the large radiation pressure in such
massive stars. The driving and damping zones are indicated by
correspondingly positive and negative values of d$W$/d$M$, where $W$
is the work integral (Cox 1980; Unno et al.~1989).  This is
illustrated in \Fig{dWdm} for the 300\,\msol star. Note that the outer
damping regions (see \Fig[a]{dWdm}) give a negligible contribution to
the work integral due to the very small amount of mass involved (see
\Fig[b]{dWdm}).  The same behavior of the work integral is also found
for the other masses studied.

\ifthenelse{\boolean{emul}}{
\vspace{1.5\baselineskip}
\noindent
\includegraphics[width=\columnwidth]{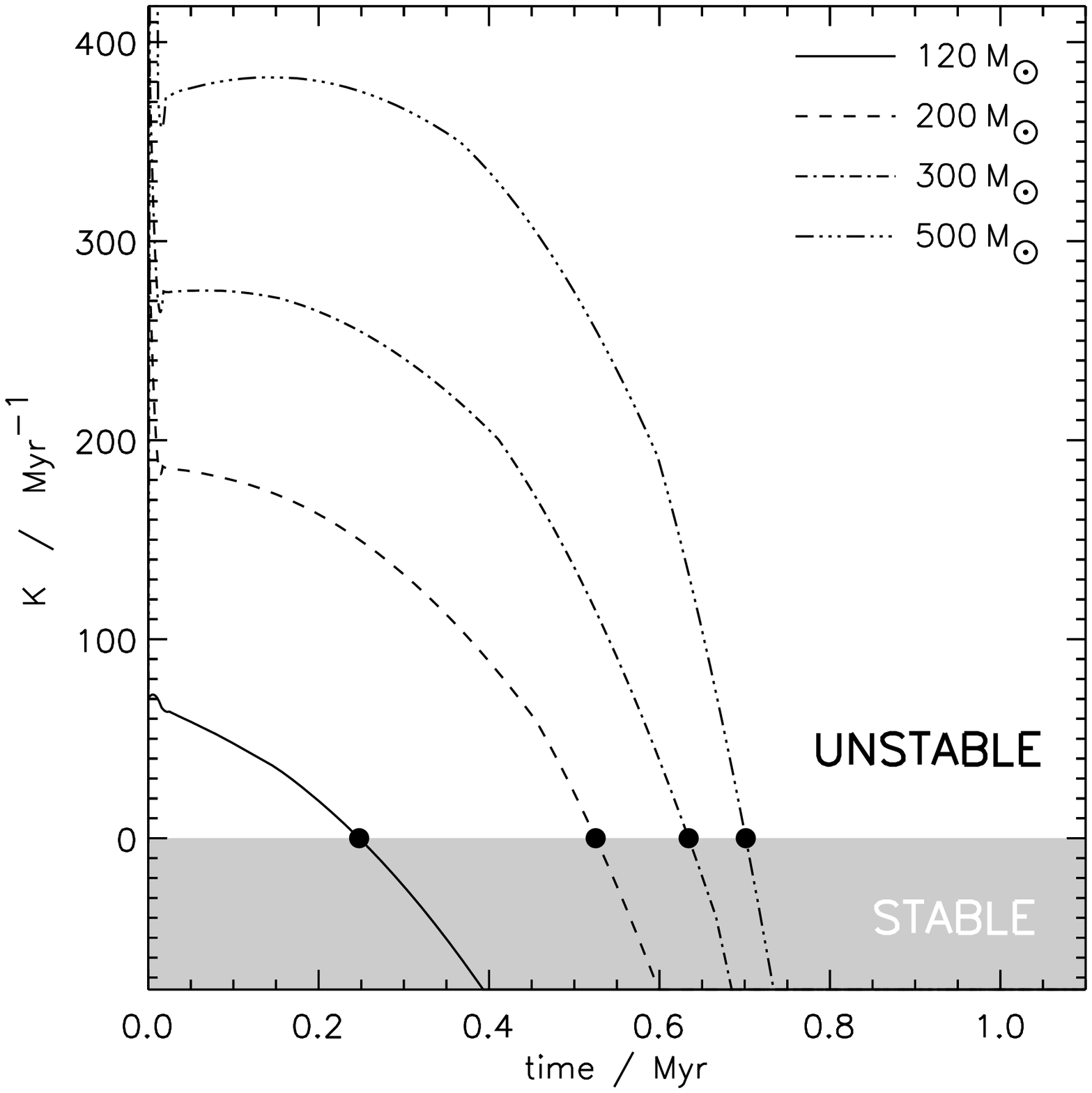}
\figcaption{\FigGrowthRate}
\vspace{1.5\baselineskip}}{}

\subsubsection{Effects of evolution}
\lSect{evol}

It is known from previous studies of solar metallicity stars that high
mass stars eventually stabilize as they evolve away from the ZAMS
(Schwarzschild \& H\"arm 1959; Stothers and Simon 1968; Aizenman et
al.~1975; Maeder 1985). Stabilization comes about because of the
increase of the mean molecular weight in the inner regions which increases
the central condensation (see, \textit{e.g.}, Maeder 1985 for details).  The same
properties are also found here for metal-free stars. As they evolve away
from the ZAMS, their $e$-folding time increases, indicating a decrease
of the driving term due to nuclear energy compared to the damping term
from heat leakage in the envelope.  This is a direct consequence of
the increase of the central density relative to the mean density.  The
stars become stable (negative $K$) on the main sequence after
a decrease of the central hydrogen mass fraction.  This is illustrated
in \Fig{K}, which displays the stability coefficient, $K$, as a function
of time for different masses, and in \Fig{A}, which shows the time
integration of $K$ during core hydrogen burning.  
In Fig. 4,  the larger value
of $K$ at early times and its rapid variation with time are
 due to the adjustment of the stellar structure when the
initial CNO is produced (see \S 2).
The results are also
summarized for the different models in Table 2 which gives the number
of $e$-foldings, \Amax, by which the amplitude of an initial
perturbation would grow in the framework of linear stability analysis,
as derived from integration of the stability coefficient with time
(see also \Fig{A}).

\ifthenelse{\boolean{emul}}{
\vspace{1.5\baselineskip}
\noindent
\includegraphics[width=\columnwidth]{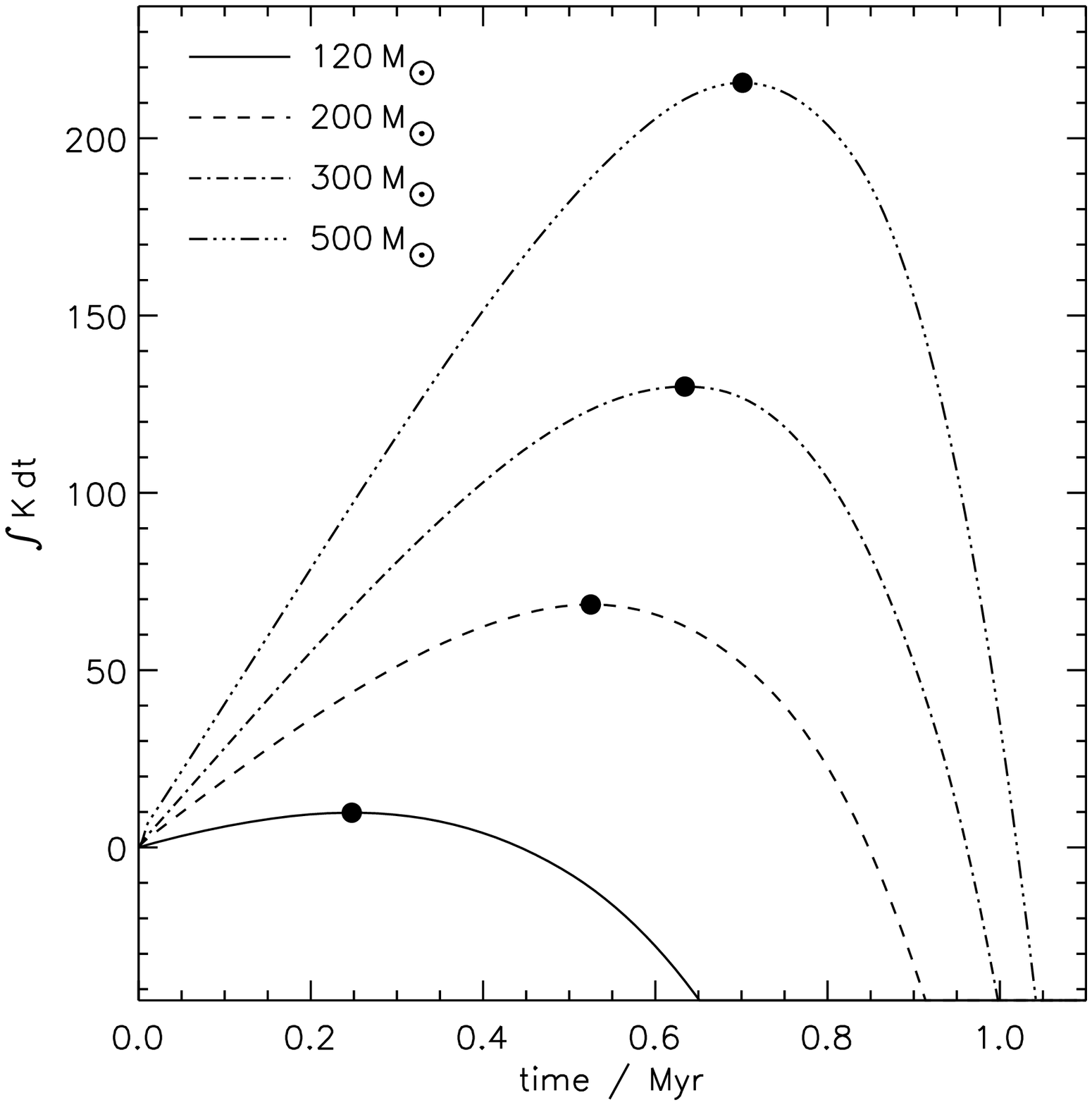}
\figcaption{\FigAmplitude}
\vspace{1.5\baselineskip}}{}

After reaching pulsational stability during central hydrogen burning,
all stars studied remained stable for the rest of central hydrogen
burning and most of central helium burning as well.  Stars of
$\lesssim200\,\Msun$ remained stable throughout core helium burning,
but stars of initial mass $\simgr300\,\Msun$ became unstable
again towards the end of helium burning.  This resurgence of
instability happens for a central helium mass fraction $\Yc = 0.1$ in
the 300\,\msol star and $\Yc=0.2$ for the 500\,\msol star,
corresponding to 30 and 50\,\kyr before the end of central helium
burning, respectively.  
The unstable mode
is the fundamental mode, characterized by $P = 14.2\,\Day$ and $\taud
= 0.13\,\yr$ for the 300\,\Msun star and $P = 40\,\Day$ and $\taud =
0.16\,\yr$ for 500\,\Msun star.  The pulsations in these blue
supergiants, with $\Teff \sim \Ep4$ K,
exist only in a surface layer that contains very little
mass.
The 500\,\Msun star, however, expands further and becomes a red
supergiant with an extended convective envelope that is characterized
by increasing period and decreasing $e$-folding time. 


The instabilities during core helium burning are characterized by
small values of \taud compared to the pulsation period.  This
indicates the presence of a strong instability that can develop rapidly
and reach large amplitudes.  Inspection of the work integral
shows that the driving zone is located in the hydrogen/helium
ionization zone and the excitation mechanism is the standard
$\kappa$-mechanism.

Note also that the helium cores would quickly become unstable to the
$\epsilon$-instability, just like their higher metallicity
counterparts, if the hydrogen envelope were removed, \textsl{e.g.}, to
a binary companion.

\subsection{Effect of $Z$} 
\lSect{Z}

\Tab{ZAMS} gives the pulsational properties of ZAMS models 
as a function of $Z$.
All models analyzed 
for $Z \, > \, 0$ are vibrationally unstable.
 The fundamental mode is the pulsation mode and
the overtones are stable.  Only for 120\,\msol with $Z \, \ge \,
 0.01$,
 are both the fundamental (F) and the first overtone (1H) unstable,
with an $e$-folding time for 1H much shorter than for the F mode.
\Tab{ZAMS} also gives the results for solar metallicity models
calculated with pure electron scattering opacity.  For such models,
our results are in good agreement with previous results based on pure
electron scattering or on simple opacity laws.  Schwarzschild \& H\"arm
(1959) find for 121\,\Msun, $P = 0.38\,\Day$ and $\taud =
1800\,\yr$, and for 218\,\Msun, $P = 0.54\,\Day$ and $\taud =
930\,\yr$.  Aizenman et al.~(1975) obtain for 110\,\Msun, $P =
0.37\,\Day$ and $\taud = 644\,\yr$.

Inspection of \Tab{ZAMS} shows that for a given mass, the period
decreases with metallicity.  This is a consequence of the
smaller radius and more compact structure of metal-poor stars when
they reach thermal equilibrium on the ZAMS, a well known effect of $Z$
on evolutionary models (\textsl{e.g.}, Ezer \& Cameron 1971;
Baraffe \& El Eid 1991; Meynet et al 1994).  The period, $P$, is 
related to the dynamical time scale of the star, $P \propto
\sqrt{R^3/M}$.  Another important result of \Tab{ZAMS} is the increase
of \taud as $Z$ decreases from $Z = 0.02$ to 0.  This indicates that
metal-free stars, although unstable, require a much longer time for
the growth of the oscillation amplitudes than solar metallicity stars.

We also note a significant difference between \taud obtained for $Z =
0.02$ models calculated using the most recent opacities, and the
electron scattering cases.  This highlights an important contribution
of opacity to the instability of very massive stars with solar
metallicity. Indeed, the driving zones in the $Z = 0.02$ ZAMS models
with $M \ge 200 \Msun$ are located at $\log T \sim 5.2 - 5.3\,\K$ and are
associated with a sharp peak of the opacity in this region due to the
contribution of heavy elements (Iglesias \& Rogers 1996).  For
120\,\Msun with $Z \ge 0.01$, our results are in agreement with the
analysis of Glatzel \& Kiriakidis (1993): the driving for the first
overtone is due to the $\kappa$-mechanism whereas for the fundamental
mode, both the $\epsilon$- and the $\kappa$- mechanisms contribute to
the driving.

As $Z$ decreases, we find a transition between nuclear-driven
pulsation and opacity-driven pulsation (\textsl{i.e.}, the $\epsilon$-
and $\kappa$-mechanisms). The $\epsilon$-mechanism is dominant for
$Z\,\simle \, 2\E{-4}$ and the $\kappa$-mechanism dominates for $Z \,
\simgr \, 2\E{-3}$, for the mass range of interest.  The increase of
\taud as $Z$ decreases for $Z \, \simle \, 2\E{-4}$ is partly due to
the different temperature dependence of the nuclear energy generation
from the CNO cycle.  As shown in \Tab{ZAMS}, central hydrogen burning
proceeds at higher $T$ for decreasing $Z$.

Locally, the temperature dependence of the specific nuclear energy
generation rate, $\enuc$, depends on temperature as $\enuc \propto
T^{\nu}$, where $\nu := \partial \ln \enuc / \partial \ln T$.  Since
the temperature sensitivity, $\nu$, of hydrogen burning by the
CNO-cycle decreases with increasing temperature (\textsl{e.g.},
Kippenhahn \& Weigert, 1992, Figure~18.8) the metal-rich stars have a
stronger sensitivity than the metal poor stars.  We find $\nu \sim 10$
for $Z = 0$ and $\nu$ increases with $Z$ up to $\sim 14$ for $Z =
0.02$ in the 120\,\Msun star.
Analytically $\nu$ can be computed from $\nu = (\tau-2)/3$ where 
\begin{equation}
\tau
= 4.248 \left(Z_1^2 Z_2^2 \frac{A_1 A_2}{A_1+A_2}/T_9\right)^{1/3}
\;.
\end{equation}
$A_1$, $A_2$, $Z_1$, $Z_2$ are mass number and charge number of the
two input nuclei, and $T_9$ is the temperature in \Ep9\,\K (Clayton
1968).  For \I{14}{N}(p,$\gamma$), $\nu=14$ at $T_9 = 0.04$ and
$\nu=10$ at $T_9=0.1$, in agreement with the values obtained above.

Since the driving of pulsations due to the $\epsilon$-mechanism depends
directly on $\nu$ with $W \propto \nu$, it is larger in metal-rich
than in metal-free stars, contributing to the decrease of $\taud$ =
$1/K \, \propto \, 1/W$ as $Z$ increases.  Another contribution to
$\taud$ comes from the total pulsation energy $\EP$ $\propto$ $1/P^2$,
since $ K \, \propto \, 1/\EP \, \propto P^2$ (see next section
\Eqs{Lpuls} and \Eqff{EP}) and $P$ increases with $Z$, as mentioned above.
As long as the contribution of metals to the opacity is negligible,
this explains why $\taud = 1/K \propto 1/P^2W$ decreases as $Z$
increases.  Once the $\kappa$-mechanism operates in the envelope, it
yields a dramatic increase of $W$ and significantly smaller \taud, by
up to six orders of magnitude (see \Tab{ZAMS}).

\begin{table*}
{\centering
\caption{%
Properties of ZAMS models as a function of mass and metallicity.
\lTab{ZAMS}}
\begin{tabular}{rlrrrll}
\tableline
\tableline
{$M$ (\msol)}  & {$Z$} & {\te (\kK)} & 
{$R$ (\Rsun)} & {\Tc (\Ep8\,\K)} & {$P$ (\Day)} &
{\taud (\yr)} \\
\tableline
120 & 0                   &  99.2 &  4.6 & 1.38 & 0.064       & 1.4\E4\\
    & 2\E{-6}             &  72.9 & 8.2 & 0.74 & 0.16        & 5.6\E3\\
    & 2\E{-5}             &  67.1 & 9.6 & 0.63 & 0.20        & 4.2\E3\\    
    & 2\E{-4}             &  62.6 & 11.0 & 0.55 & 0.25        & 3.7\E3\\
    & 0.01                &  53.3 & 15.2 & 0.45 & 0.34 (0.16) & 4.6\E3 (2.3\E{1}) \\
    & 0.02                &  48.9 & 18.4 & 0.44 & 0.36 (0.26) & 1.9\E2 (1.4\E{-3}) \\
    & 0.02 ($e^-$ scatt.) &  67.1 & 13.2 & 0.45 & 0.34        & 7.8\E2             \\ 
\tableline
200 & 0                   & 104.2 &  6.0 & 1.43 & 0.083       & 3.7\E3             \\
    & 0.02                &  44.9 & 32.1 & 0.46 & 0.76        & 1.3\E{-3}          \\
    & 0.02 ($e^-$ scatt.) &  70.5 & 17.5 & 0.47 & 0.45        & 5.3\E2             \\
\tableline
300 & 0                   & 107.0 &  7.6 & 1.47 & 0.10        & 2.6\E3             \\
    & 2\E{-4}             &  68.1 & 18.2 & 0.60 & 0.39        & 1.3\E3             \\
    & 0.02                &  39.8 & 54.1 & 0.47 & 1.79        & 2.1\E{-3}          \\
    & 0.02 ($e^-$ scatt.) &  72.5 & 22.0 & 0.48 & 0.55        & 4.3\E2             \\
\tableline
500 & 0                   & 109.2 & 10.0 & 1.51 & 0.13        & 2.0\E3               \\
\tableline
\end{tabular}
\\[0.5\baselineskip]} \textsc{Note.}---
The pulsation period and $e$-folding time are given for the
fundamental mode. Only for 120\,\msol with $Z \ge 0.01$ the fundamental
mode and first overtone (in parentheses) are both unstable. For $Z =  
0.02$, results based on models calculated with pure electron
scattering opacity ($e^-$ scatt.) are also given.  
\end{table*}

\ifthenelse{\boolean{emul}}{
\begin{table*}
}{
\begin{table}
}
{\centering
\caption{Properties of $Z=0$ stars during core hydrogen burning.  \lTab{stab}}
\begin{tabular}{rrrrr}
\tableline
\tableline
$M$ (\Msun) & \tH (\Myr) & \tS (\Myr) & $Y_c$ (\%) & \Amax \\
\tableline
120   & 2.35 & 0.25 & 69 &   10 \\
200   & 2.15 & 0.52 & 58 &  69 \\
300   & 1.94 & 0.63 & 51 & 130 \\
500   & 1.74 & 0.70 & 46 & 216 \\
\tableline
120es & 1.74 & 0.97 & 33 & 1091 \\
\tableline
\end{tabular}
\\[0.5\baselineskip]} \textsc{Note.}---
The table gives the initial mass $M$, the total
hydrogen-burning lifetime of the star \tH, the age at which the
star becomes stable \tS,
the central hydrogen mass fraction at
this time $Y_c$ (initial value is 76\,\%),
 and the  number of $e$-folds an initial pulsation has
grown since the ZAMS by then, $\Amax=\int_0^{\tS}K(t)$d$t$.
\ifthenelse{\boolean{emul}}{
\end{table*}
}{
\end{table}
}

\ifthenelse{\boolean{emul}}{
\begin{table*}
}{
\begin{table}
}
{\centering
\caption{Mass loss estimates of stars at ZAMS.  \lTab{stru}}
\begin{tabular}{rlrrr@{}l}
\tableline
\tableline
$M$ (\Msun) & 
$Z$ &
$\EP (10^{48}$ erg) & 
\Mdot (\Msunyr) & 
\multicolumn{2}{c}{\DM (\Msun)} \\
\tableline
120  & 0 &  0.5 & 7.5\E{-7} &  0.2 \\
200  & 0 &  1 & 5  \E{-6} &  2.5&   \\
300  & 0 &  2.5 & 1.3 \E{-5} & 8.5&   \\
500  & 0 & 6 & 3.5 \E{-5} & 24.5&   \\
\tableline
120    & 0.02 (e$^-$ scatt.) & 0.2 & 1.7 \E{-5} & 17& \\
\tableline
\end{tabular}
\\[0.5\baselineskip]} \textsc{Note.}---
The table gives the initial mass, $M$, metallicity, $Z$, pulsation
energy, \EP, mass loss rate at the ZAMS, and an estimate of the total
mass lost during the time the stars are unstable (\tS, see \Tab{stab})
during central hydrogen burning, \textsl{i.e.}, $\DM = \Mdot \tS$.
The results for 120\,\Msun, $Z=0.02$ and pure electron scattering
opacity is also given.
\ifthenelse{\boolean{emul}}{
\end{table*}
}{
\end{table}
}


\section{Mass loss rates}
\lSect{Mdot}

In order to derive a realistic mass loss rate, non-linear calculations
would be required to determine the pulsation amplitudes and
dissipation due to shock wave formation.  Non-linear calculations have
been performed for the solar metallicity case by various authors
(Appenzeller 1970ab; Ziebarth 1970; Papaloizou 1973a; Talbot 1971ab)
who, unfortunately, give different results.  Appenzeller (1970ab)
finds an upper limit $\Mdot=5\E{-5}\,\Msunyr$ for a 130\,\Msun star
and $\Mdot=5\E{-4}\,\Msunyr$ for a 270\,\Msun star, whereas Papaloizou
(1973b) finds stronger dissipation effects and mass loss rates less
than \Ep{-6}\,\Msunyr for similar masses.

Since the dissipation effects in Appenzeller (1970ab) may be
underestimated compared to other studies, yielding larger mass loss
rates, we used his results to derive an upper limit for \Mdot for the
$Z=0$ models.  Appenzeller found that when the velocity
amplitude at the surface approached the sound speed, mass loss sets
in.  An upper limit of the mass loss rate came from assuming that
the entire rate of gain of pulsation energy, \Lpuls, was used to
accelerate the matter to escape velocity, $\vesc=\sqrt{2GM/R}$.
Thus one has:
\begin{equation}
\frac{\Mdot}{2}\vesc^2=\Lpuls
\;.\lEq{Lcmp}
\end{equation}

\Lpuls can be estimated from the total pulsation energy \EP,
averaged over one period, and
the stability coefficient, $K$ (see 
Cox 1980), with 
\begin{equation}
\Lpuls = 2K \EP
\;,\lEq{Lpuls}
\end{equation}
where
\begin{equation}
\EP=\frac{1}{2} \left(\frac{2\pi}{P}\right)^{\!2}\int_0^M\dr^{\!2} \, dM
\;.\lEq{EP}
\end{equation}

Since the relative pulsation amplitude, $\dr/\drs$, is given by our
linear stability analysis, \drs and thus \EP are estimated
by assuming, like Appenzeller (1970ab), that the
surface velocity amplitude reaches the sound speed, \csound,
\textsl{i.e.}:
\begin{equation}
\frac{2\pi}{P} \drs = \csound
\;.\lEq{vsurf}
\end{equation}

For the zero metallicity stars from 120 to 500\,\Msun, \csound is
typically $(1-1.6)\E7\,\cms$.  The resulting energy, \EP, and mass
loss rates for the $Z=0$ models on the ZAMS are given in
{\Tab{stru}}. Results are also given for the 120\,\Msun star with
$Z=0.02$ and pure electron scattering opacity.  Our upper limit for
\Mdot in this case is in good agreement with Appenzeller's value
of 5\E{-5}\,\Msunyr for similar masses.  A crude upper limit for the
total mass lost \DM during hydrogen burning is derived assuming that
mass loss proceeds at constant rate as long as the star remains
unstable, \textsl{i.e.}, during the time \tS (see \Tab{stru}).  In any
case, \DM does not exceed 5\,\% of the total mass for zero-metal
stars.  Note that this estimate neglects the time required for the
pulsation amplitude to grow to the value \drs characterizing the
beginning of mass loss (\Eq{vsurf}).  Given the non negligible value of
\taud compared to \tS (see \Tab{ZAMS}), this time may be significant for
the 120\,\Msun star.  In fact, this star may never encounter
pulsationally driven mass loss.  Moreover, since the stars stabilize
as soon as they evolve away from the ZAMS (see \Sect{evol}) with
decreasing $K$ (see \Fig{K}), one can expect decreasing \Mdot along
the main sequence, and thus even smaller values for \DM. 
We thus find that even for
500\,\Msun star, pulsations do not lead to significant mass loss on
the main sequence.


For post-main sequence evolution,
as found in \Sect{evol}, only stars with initial masses 
of $\simgr \, 300 \,\msol$ become unstable again during helium burning
and can undergo another phase of mass loss.
Any estimate of mass loss is hampered by the lack of non-linear
calculations during these late phases of
evolution.  Moreover, as the star evolves toward low effective
temperature and eventually becomes a red supergiant, it develops a
deep convective envelope, the opposite of main sequence stars.  This
adds additional difficulty to any derivation of \Mdot since convection
can have an important contribution to the damping or driving of the
pulsation in this case.

\section{Discussion and conclusions}
\lSect{concl}

Our analysis has shown that even though metal-free stars
with $120\,\Msun \le\, M \, \le 500\,\Msun$ are radially unstable on
the main sequence (like their metal-enriched counterparts), the
efficiency of the $\epsilon$-mechanism is reduced because of the
hotter central $T$ and more compact structure.  Consequently, the
$e$-folding times characterizing the growth of amplitudes are much
longer.  After burning some hydrogen, the stars stabilize
(\Tab{stab}).

In order to know the effects of such oscillations and accurately
estimate the resulting pulsationally driven wind, non-linear
calculations would be required.  We have not done those and have only
estimated the mass loss using simple arguments based upon previous
non-linear calculations.  Adopting the approach of Appenzeller
(1970ab) that the mass loss can be derived from the energy gain rate
of the pulsation when its surface velocity reaches the local velocity
of sound at least gives a quantitative result which is a reasonable upper
limit (\Sect{Mdot}).  For the 120\,\Msun star we find negligible mass
loss during the central hydrogen burning phase.  The integrated mass loss
during central hydrogen burning of about 2.5\,\Msun for a 200\,\Msun
star increases to about 25\,\Msun for the 500\,\Msun star - less than
5\,\% of its initial mass.

We also note that Papaloizou (1973b) disagrees with the results of
Appenzeller (1970a) and finds an important limitation of the pulsation
amplitude, which cannot drive any mass loss. If this is the case for
the solar metallicity stars, it is even more so for metal-free stars.
Therefore we conclude that mass loss during central hydrogen burning
remains unimportant for the evolution of all the Z=0 stars
investigated (\Tab{stru}).

Finally, it is worth noting that the strong instability found at the
end of core helium burning for $M \ge 300\,\msol$ could have important
consequences.  What exactly happens depends on the remaining time
before collapse (only a few 10\,\kyr for the 300 and 500\,\Msun stars
investigated here) and, especially, on the strength of the mass loss
rates achieved in these pulsating giant stars with convective
envelopes.  This is an important issue regarding the gamma-ray burst
scenario suggested by Fryer et al.~(2000) and needs further
investigation.

From our estimates of the mass loss rates we conclude that primordial
very massive stars of $\lesssim500\,\Msun$ (but over $100\,\Msun$)
reach core collapse with helium cores massive enough to encounter the
pair creation instability. Single stars probably do not lose their
hydrogen envelope for initial masses of $\lesssim300\,\Msun$.  The
nucleosynthetic yields of these stars can be important for the
chemical evolution of the early universe, as they may eject up to
$60\,\Msun$ of \I{56}{Ni} for a $\approx250\,\Msun$ star (Heger \&
Woosley 2000). Still heavier stars make black holes (Bond, Arnett \&
Carr 1984).  Extrapolating from our results, we speculate that stars
of $\simgr1000\,\Msun$ might experience such strong mass loss that
they end their lives as smaller objects.  But clearly the stability of
these massive ``first stars'' is a subject worthy of additional
exploration beyond the first linear stability analysis presented here.

\acknowledgements

We are grateful to Gilles Chabrier for many valuable discussion.  IB
thanks the Astronomy Departments of University of California at
Berkeley and Santa-Cruz for hospitality during the completion of this
work.  This research was supported, in part, by Prime Contract
No.~W-7405-ENG-48 between The Regents of the University of California
and the United States Department of Energy, the National Science
Foundation (AST 97-31569, INT-9726315), and the Alexander von
Humboldt-Stiftung (FLF-1065004).

{}


\ifthenelse{\boolean{emul}}{}{

\clearpage
\onecolumn

\ifthenelse{\boolean{\IncludeFigures}}{
\renewcommand{\figcaption}[2][]{
\clearpage
\begin{figure}
\epsscale{0.8}
\plotone{#1}
\caption{#2}
\end{figure}
}}{}

\figcaption[\FigHRDmassFile]{\FigHRDmass}
\figcaption[\FigHRDZFile]{\FigHRDZ}
\figcaption[\FigWorkFileA]{\FigWork}
\figcaption[\FigGrowthRateFile]{\FigGrowthRate}
\figcaption[\FigAmplitudeFile]{\FigAmplitude}
}

\end{document}